# Cortical Brain Computer Interface for Closed-Loop Deep Brain Stimulation

Jeffrey A. Herron Ph.D., Margaret C. Thompson, Timothy Brown,
Howard J. Chizeck, Sc.D., Jeffrey G. Ojemann, M.D., and Andrew L. Ko, M.D.

*Abstract*— Essential Tremor is the most common neurological movement disorder. This progressive disease causes uncontrollable rhythmic motions—most often affecting the patient's dominant upper extremity—that occur during volitional movement and make it difficult for the patient to perform everyday tasks. Medication may also become ineffective as the disorder progresses. For many patients, deep brain stimulation (DBS) of the thalamus is an effective means of treating this condition when medication fails. In current use, however, clinicians set the patient's stimulator to apply stimulation at all times—whether it is needed or not. This practice leads to excess power use, and more rapid depletion of batteries that require surgical replacement. In the work described here, for the first time, neural sensing of movement (using chronically-implanted cortical electrodes) is used to enable or disable stimulation for tremor. Therapeutic stimulation is delivered only when the patient is actively using their effected limb, thereby reducing the total stimulation applied, and potentially extending the lifetime of surgically-implanted batteries. This work, which involves both implanted and external subsystems, paves the way for the future fully-implanted closed-loop deep brain stimulators.

## I. INTRODUCTION

ESSENTIAL Tremor (ET) is the most common neurological movement disorder [1] and it can have a dramatic impact on patient quality of life [2]. ET causes uncontrollable rhythmic motions—most often affecting the patient's dominant upper extremity—during volitional movement. These uncontrollable movements make it difficult to perform everyday tasks such as eating, drinking, writing, or other activities that require fine motor control. While pharmaceutical agents (e.g., propranolol) can help relieve symptoms for some individuals, medication is often ineffective or poorly tolerated in the long run: ET is a progressive disease, and its symptoms worsen over time [3]. Deep brain stimulation (DBS) of the thalamus is often used to give otherwise unresponsive patients control of their limb and freedom from disabling tremor symptoms [4,5].

Current DBS treatment of ET is performed using a battery-powered, implantable pulse generator (IPG) surgically implanted in the chest. This IPG then sends electrical stimulation to a lead inserted through a hole drilled in the skull to electrical contacts in the ventral intermediate (VIM) nucleus of the thalamus [5]. While the underlying causes of ET in the brain are unknown, and the means by which thalamic stimulation treats ET are unknown [6], electrical stimulation of the VIM can be remarkably effective for reducing tremor in ET patients [4,5].

However, DBS for ET comes with several drawbacks. Stimulation can cause unpleasant side-effects for patients, such as paresthesias (abnormal facial or extremity sensations like tingling), dysarthria (speech impairment), and ataxia (dyscoordination) [7]. Additionally, most IPGs are non-rechargeable and require surgical replacement when the battery is depleted. Current clinically- and FDA-approved DBS systems run in an "open-loop" manner, meaning that they deliver stimulation at a steady rate for as long as the device is running. Changes to this therapy require manual intervention by a clinician to adjust the amplitude, pulse width, and frequency of the stimulation waveform. This "open-loop" means of applying stimulation can lead to additional problems for the patient. Patient symptoms and side effects vary widely and are difficult to evoke consistently during bedside testing, and so it can be difficult to determine appropriate stimulation parameters for each patient [8]. The current practice of applying open loop stimulation, even at times when the stimulation is not necessary, is wasteful and can result in more frequent battery replacement surgeries or larger implanted devices than would otherwise be needed.

This work is supported by a donation from Medtronic and by Award Number EEC-1028725 from the National Science Foundation for the Center for Sensorimotor Neural Engineering. The content is solely the responsibility of the authors and does not necessarily represent the official views of the National Science Foundation or Medtronic. This research was also supported by the Department of Defense (DoD) through the National Defense Science & Engineering Graduate Fellowship (NDSEG) Program.

J. Herron was with the Dept. of Electrical Engineering at the University of Washington, Seattle, WA when this work occurred. After completing his PhD, he has since joined Medtronic Inc.
M. Thompson, and H. Chizeck are with the Department of Electrical Engineering, T. Brown the Department of Philosophy, J. Ojemann and A. Ko with the Department of Neurological Surgery, all University of Washington, Seattle, WA. For all material correspondence, please use chizeck@uw.edu.

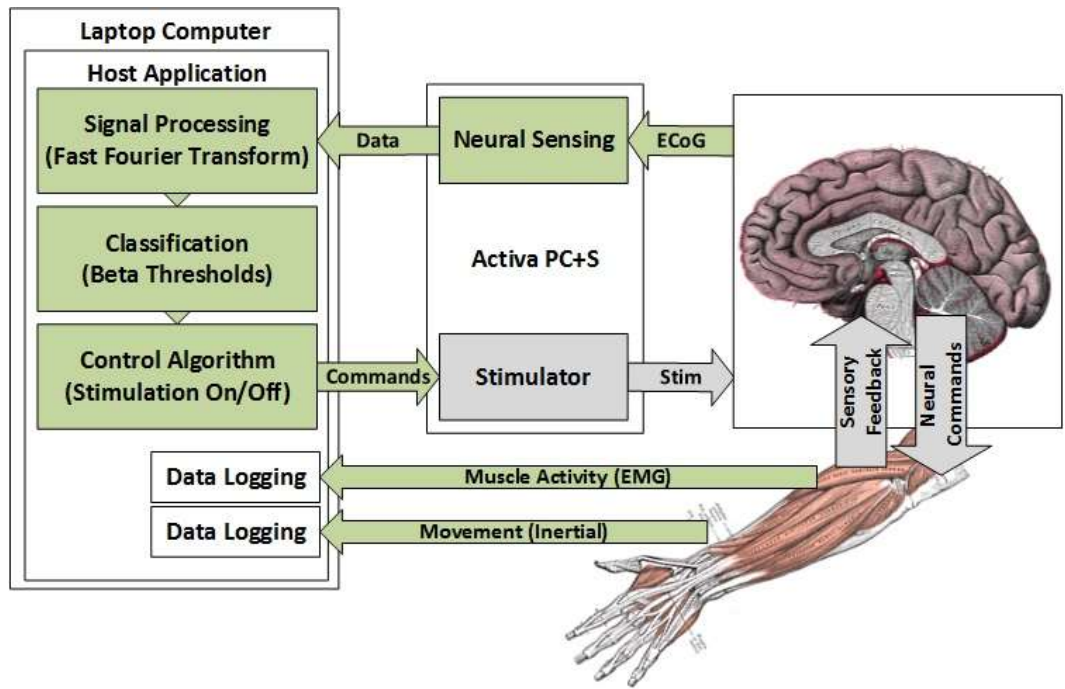

Fig 1: Block diagram illustrating the experimental setup for our closed-loop DBS experiments. ECoG data, sensed by the Activa PC+S and streamed by the Nexus-D, is processed and classified on a laptop computer. When the Beta band-power crosses thresholds defined for movement and rest, the stimulation is turned on or off. Limb inertial and EMG recordings are logged and timestamped separate from the closed-loop.

"Closed-loop" methods of applying stimulation, where IPGs use sensors to collect sensor data from the patient and automatically make stimulation adjustments as needed, are being investigated as a solution to these drawbacks. For example, in the case of many ET patients, a system that only provides stimulation when the patient is moving the affected limb may be capable of treating tremor while preserving battery life and reducing side-effects. Prior applications of closed-loop DBS for ET in human subjects have exclusively used wearable inertial sensors [9], or electromyography (EMG) to provide feedback for closed-loop DBS experiments [9,10].

While wearable sensors are appropriate for use in an experimental setting, there are significant barriers to translating wearables to a clinically deployable system. Externally sensed data and externally computed control actions must be transmitted to the IPG. The computations necessary are time-intensive and introduce latency into the system. Further, the telemetry required to maintain a wireless communication channel between the IPG and external sensors is power intensive, and it is possible that the power lost on wireless telemetry would negate much of the power savings that closed-loop stimulation would deliver. Additionally, patient acceptance may be a concern, as wearing a sensor for much of the day may be uncomfortable, cumbersome, or draw unwanted attention. One possible way to address these potential problems with wearable closed-loop systems is to use signals collected from an additional set of implanted electrodes [11]. This would allow an implanted IPG to collect the data required to make control decisions without the need for worn sensors, off-board computations, or telemetry.

In this paper, we demonstrate the first use of chronic electrocorticography (ECoG) to modulate an ET patient's DBS in real-time through the use of movement-related cortical signals. This closed-loop system uses an investigational IPG with neural sensing capabilities to modulate stimulation amplitude based on motor cortex ECoG. We show that cortical movement-related beta-band desynchronization (during overt limb movement of the tremor-affected arm) can be used effectively to trigger DBS in order to reduce tremor. We demonstrate the efficacy of this system while the patient performs standard drawing tasks that are a component of a widely-accepted clinical tremor assessment as well as a prompted movement task where the patient performs movements that predictably cause tremor. Our system successfully and accurately enables and disables stimulation during periods of movement and rest in these trials—and, as a result, the system reduces the power consumed by the IPG. That is, the system applies therapeutic stimulation to treat tremor only when the patient is actively using the limb, thereby reducing the total amount of stimulation applied, and consequently may extend the lifetime of surgically-implanted batteries. This work paves the way for the future fully-implanted closed-loop deep brain stimulators for use as therapeutic tools. It also facilitates investigation of closed-loop patient-device system dynamics.

## II. METHODS

The experimental protocol used was approved by the Institutional Review Board at University of Washington Medical Center, and use of the Activa PC+S and Nexus D system was approved by the FDA through an Investigational

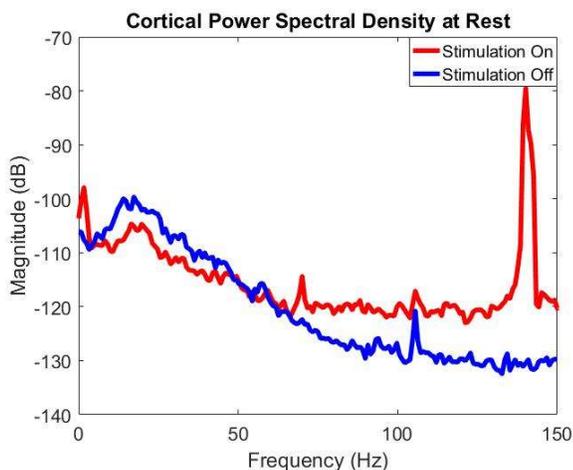

Fig 2: Cortical power-spectral density (PSD) while the patient is at rest in both stim on and stim off states. Note "stim on" results not only a large stimulation artifact at 140Hz, but also a "flattening" of the PSD with lower beta-band and higher gamma band activity. These PSD plots were calculated offline after the experiment using 20 seconds of continuous patient resting data in each state.

Device Exemption. The experiments were performed in accordance with all relevant guidelines and regulations. The subject provided informed consent according to direction from the Institutional Review Board prior to enrollment in our study.

*A. Patient and surgical procedure*

The patient is a 58-year-old right-handed man who experiences tremor during volitional movement of his right arm and leg. The patient consented to participate in our study to implant an investigational IPG developed by Medtronic, the Activa PC+S. This IPG has the capability to sense through implanted electrodes as well as provide traditional clinical therapeutic stimulation. The sensed data can be streamed to an external computational device, such as a laptop computer, through the use of the Nexus-D, another investigational device developed by Medtronic [12].

In addition to the unilateral DBS lead implanted in the VIM for therapy, we implanted a single cortical strip of electrodes (the Resume II lead) on the surface of the brain overlying motor cortex corresponding to the patient's right hand and connected directly to the Activa PC+S. Motor cortex was localized using anatomical landmarks on preoperative MRI by creating an additional plan on the intraoperative navigation system targeting the "hand knob" in motor cortex. During surgery, the strip was advanced toward this target, with placement confirmed using intraoperative CT and adjusted to place two electrodes over the hand motor cortex and two electrodes over the sensory cortex. This strip of four electrodes allows for the chronic recording of neural signals from the cortex that have been shown to correlate with hand movement. Prior work with this patient demonstrated that limb movement may be detected using the Activa PC+S based on the onset of a decrease in beta power over sensorimotor cortex by sensing differentially across M1 to S1 [9]. The specific electrodes chosen was the pair with the highest beta-band power while the patient was at rest, determined through a standardized montage-sweep provided by the Activa PC+S instruments. We also previously demonstrated the use of surface electromyography (EMG) from this patient's arm muscles to trigger stimulation during movement. Using EMG to trigger stimulation only during muscle activation resulted in both power savings and suppression of tremor [9]. The innovation of the work reported here is the use of cortical signals to directly drive the closed-loop control system.

*B. System Description*

The overall experimental system is represented in the block diagram shown in Figure 1, which outlines each component of this first neural closed-loop DBS system for ET. The Activa PC+S digitizes neural signals at a rate of 422 samples per second. The Activa PC+S then streams the data to a laptop using the Nexus-D. Each packet contains 400ms (168 samples) of data. The laptop converts the time-domain data into the frequency domain, performing a 256-sample long FFT, with a Hann window and zero padding. We then summed the FFT's discrete frequency output bins that overlap with the beta-band in order to estimate the patient's beta band-power. For this experiment, we used the bins that correspond to the frequency range of 14.8Hz to 31.3Hz. When the patient moves, the beta-band component in the signal decreases, a well-described phenomenon known as cortical desynchronization [13]. When this beta band-power drops below a manually-tuned threshold, the system ramps

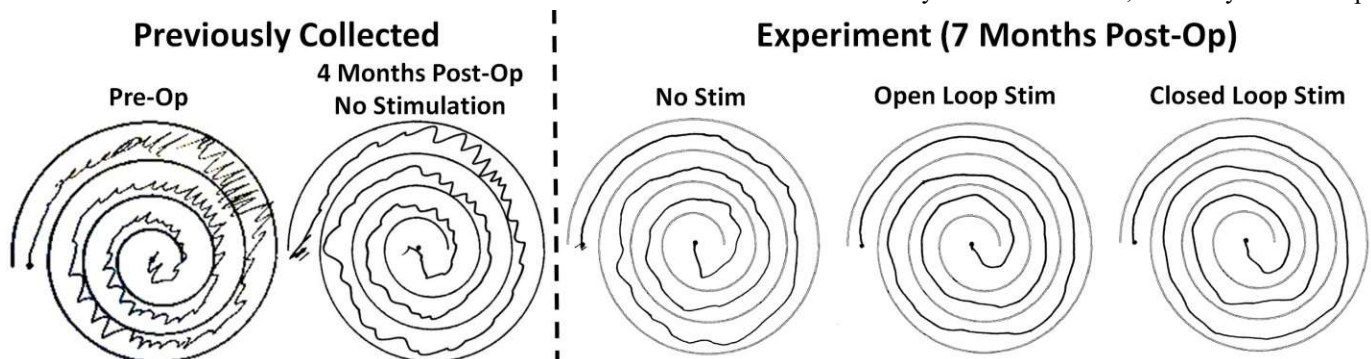

Fig 3: Patient drawn spirals collected as part of clinical tremor assessment. On the left are two spirals drawn with the dominant hand before and after DBS implant. On the right are spirals collected on the day of the experiment in each of the three stimulation states. Differences in tremor between no stimulation at four months post-op and experimental day are attributed to day to day tremor variation. Note tremor in the upper right and lower left quadrants of the spiral in the experimental no stimulation case. Comparatively there are few deviations from normal spirals in the open-loop and closed-loop cases.

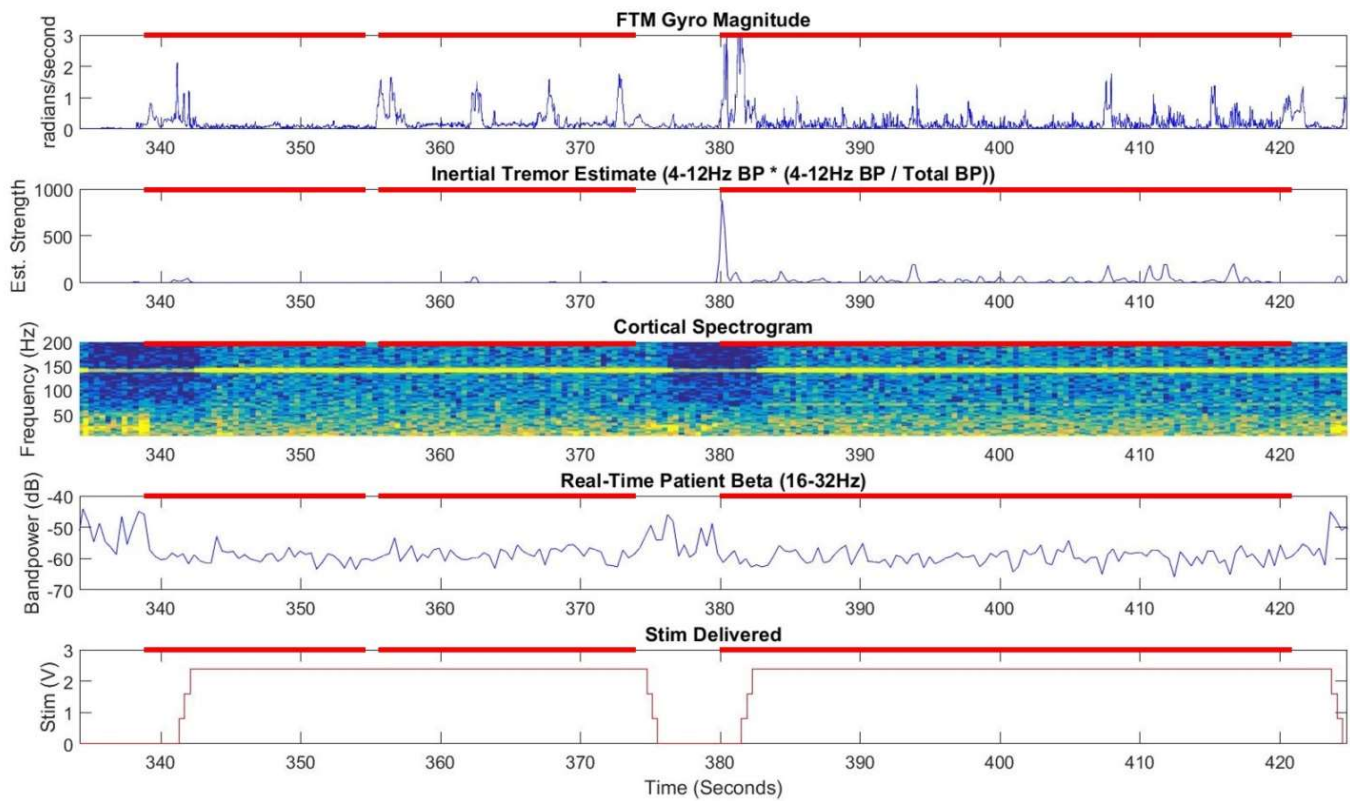

Fig 4: Task 1 results of the Fahn-Tolosa-Marin (FTM) tremor assessment task with time plots of the patient's gyroscope magnitude, tremor estimate, cortical spectrogram, beta-band power estimate, and stimulation delivered with closed-loop stimulation. Red bars indicate when patient was performing one of the three tasks. There is a break between tasks two (connecting dots) and three (writing) where researchers changed pages. Note stimulation at zero volts before tasks began, during the break, and at the end of the trial

stimulation up to the normal clinical setting, or "on" state, to deliver therapeutic stimulation and treat the tremor. Similarly, when the movement ends and the motor circuits re-synchronize, the beta band-power will rise. When it crosses a second manually-tuned threshold, the stimulation is turned off, thus conserving power when movement is not occurring.

To simplify the development of this proof of concept, we iteratively tuned the thresholds used to trigger stimulation changes. The two manually-tuned thresholds represented the movement-initiation band-power threshold and the rest initiation band-power threshold. The range over which the thresholds were defined was based on the mean beta-band measured first while the patient was at rest with stimulation on and second while the patient was moving with stimulation off. The specific values for the movement-initiation and rest-initiation thresholds were then manually tuned within this calibrated range while the patient alternated between movement and rest. For these experiments, the off threshold was calibrated to be at a higher beta band-power than the on threshold, which results in a dead-band between the two thresholds where no action is taken. We averaged the beta band-power estimation across five packets (or two seconds) in order to gain better specificity for movement detection. This averaging acts as a type of low-pass filter on the power signal and the system uses more information to make a control decision which reduces the impact of noise on the algorithm. However, averaging the feedback signal in this way introduces system delays that reduce overall system responsiveness.

There are several complicating factors that need to be considered when designing experimental code to perform this task. We noticed that there was a dramatic effect of stimulation on the patient's sensed cortical signal, as shown in Figure 2 while the patient was at rest. Turning stimulation on resulted in a "flattening" of the power-spectral density of the streamed data. While it is unclear if this flattening was due to physiology or device, the artifact was robust and repeatable. Fortunately, the magnitude of movement-related desynchronization was sufficient for use as a biomarker despite the noise in these signals and the reduction in band-power when stimulation is present. However, the fact that VIM stimulation resulted in lower cortical beta was a key rationale for the use of two separate manually-tuned thresholds instead of a single one.

### C. Experimental Tasks

To assess the experimental system, the patient performed two sets of tasks with no stimulation applied, open-loop stimulation, and closed-loop neural-triggered stimulation. To then quantify the tremor that the patient then experiences in each task or trial with the test algorithms, we performed an offline spectral analysis on the gyroscope data from the worn smartwatch. First, we summed the band-power within a low-frequency band (approx. 4-12Hz) that captures the spectral power of the rhythmic tremor movement. Then as a simple form of artifact rejection, we scale this band-power value by the percentage of the total band-power measured from the device that is within this tremor band. This reduces the impact of large cross-band inertial signals that in our trials were not indicative of tremor but instead of task design or movement initiation. This method and the rational are discussed in our prior work with this patient during closed-loop trials using wearable sensors [9].

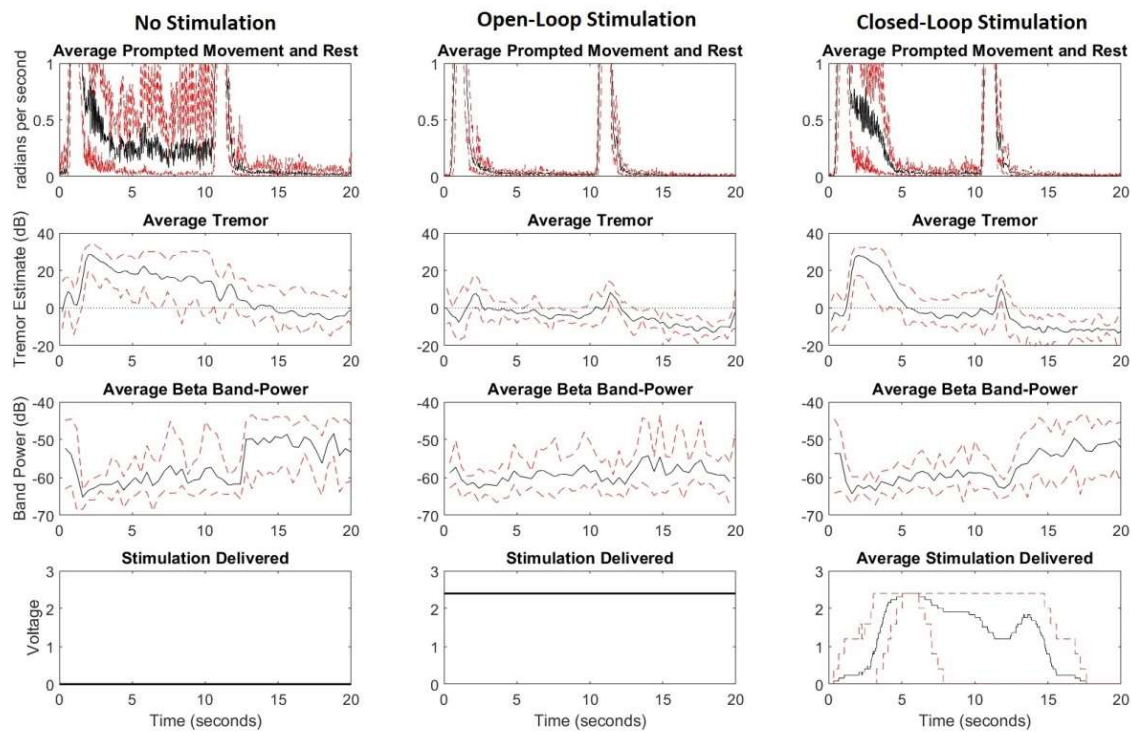

Fig 5: Task 2 results of the movement prompting task showing average gyroscope magnitude, tremor estimate, beta-band power, and stimulation delivered for each of the three stimulation configurations. For all plots, red dashed lines indicate the 0.9 to 0.1 quantile bounds. The first 10 seconds of each plot is during the prompted movement, and timespan from 10-20 seconds corresponds to the following rest period. On the left are the results from the no-stimulation case, as shown by the stimulation delivered at zero volts. In the middle are the results from the open-loop stimulation case, as shown by the 2.4V stimulation setting. On the right are the closed-loop results with dynamic stimulation driven by the cortical sensing.

The first set of experimental tasks are from the commonly-used Fahn-Tolosa-Marin tremor assessment [14], including the spiral draw, connecting dots, and sentence writing tasks. The patient was given short breaks in between the dot connecting and sentence writing tasks to give the researchers time to give the patient the next sheet of paper to work on.

For the second task to assess the experimental system, the patient was then asked to complete a series of computer-prompted tasks where he was instructed to alternate between resting his hand and bringing his hand to his mouth. This motion caused predictable tremor symptoms in this particular patient and facilitated a direct comparison of how each stimulation paradigm performed in a repeatable task. There were ten prompts each for movement and rest, with each prompt lasting for ten seconds. Each trial thus lasted a total of 200 seconds. The prompt was delivered through both visual cues presented on-screen to the patient and audio cues. While the patient was performing these tasks, we recorded data related to his arm movements with a wrist-worn accelerometer and surface EMG.

### D. Safety Considerations

Prior to closed-loop experiments, the patient's therapy settings were configured for maximal therapeutic benefit using the Medtronic clinician DBS programmer (8840) in a normal patient-programming routine. Then, using the configured electrodes and determined therapeutic amplitude as a limit, we characterized the side effect profile while increasing the stimulation amplitude ramping rate between "on" (2.4 volts) and "off" (0 volts) states. We asked the patient to report any unpleasant side-effects as stimulation was ramped up and down repeatedly with increasingly fast slew rates so we could determine how fast the closed-loop algorithm could safely ramp stimulation between the on and off states. The interval between stimulation setting updates was limited to the sensing packet rate (400ms) to avoid losing streamed data during the experiments with sensing enabled. When increasing the stimulation by 200mV every 400ms, the patient reported mild paresthesias that were not unpleasant. At 1000mV every 400ms, these sensations were more marked and surprising. Given his unpleasant reaction at the 1000mV level, we used 800mV/400ms as the ramping rate for closed-loop stimulation testing. This would result in the system being able to transition between the on and off states in a total 1.2 seconds.

## III. RESULTS

The spirals from the tremor assessment drawn during no-stimulation, open-loop stimulation, and closed-loop stimulation are shown in Figure 3. Additional spirals collected before and four months after surgery are shown for comparison. As can be seen through visual inspection, the patient's tremor on the day of the experiment was lower than previous assessments. It is well known that the severity of tremor can vary day to day, but even during this low-amplitude session there are clear hallmarks of tremor in the spiral which were apparent in comparison to the open-loop case. The closed-loop case appears to be nearly identical to the open-loop results, indicating that the closed-loop system was able to significantly reduce his tremor while writing.

All three tremor assessment writing tasks were performed for each stimulation paradigm, and the time-series plot of the closed-loop tremor assessment trial is shown in Figure 4. In the trial, the stimulation was turned off when the patient was at rest before the trials began, and during the short break in between the second and third tasks while the researchers switched pages. Stimulation was also correctly turned on for the duration of drawing spirals, connecting dots, and writing sentences.

The results from the movement-prompting task are shown in Figure 5. Each of the ten pairs of movement and rest periods were averaged together for each of the trials with the three different stimulation settings. The cumulative magnitude of the three axes of the inertial gyroscope is displayed in the top row of the figure. By analyzing the gyroscopic data, we can estimate how much tremor the patient experienced at a given moment. Due to patient reaction times and time to transition between states, initiation of movement is offset from the prompts at 0 and 10 seconds by approximately 1 second; this is indicated by the large discrete peaks that exceed the window bounds. Tremor appears as a noisy signal in-between the transient peaks, while the patient is holding his hand at his mouth. The strength of the tremor is estimated through spectral analysis to remove the influence of the transient peaks, as shown in the second row.

With no stimulation, there is tremor consistently throughout the prompted action and no tremor at rest. Open-loop stimulation results in low tremor throughout the prompted action and while at rest. The closed-loop system results in brief tremor at the start of the movement, followed by rapid reduction to the open-loop levels. In the third row, the average cortical beta-band power is shown. With stimulation off, there is a clear difference between the mean beta-band during rest and movement. Stimulation effects on the beta band make this distinction more difficult to see during open-loop stimulation.

During closed-loop stimulation, the beta-band tends to follow a more "normal" trend with low beta-band power during movement and higher beta-band power during rest. The average stimulation delivered in each trial is depicted in the final row. For the closed-loop stimulation trial, the average stimulation rose to clinical amplitudes during every movement and dropped to zero during every rest. The mean amplitude and quantile bounds reach 2.4 volts about five seconds after the prompt, indicating that for each and every movement period, stimulation reached clinical levels. The movement of the limb from the action to rest state also triggers stimulation, as can be seen in an average increase in the stimulation amplitude around 14 seconds. However, by the 18-second mark, prior to the next movement prompt, stimulation was off for all trials.

## IV. DISCUSSION

The results show that the cortical movement-related beta-band desynchronization is an effective signal to trigger DBS in order to reduce tremor. While the patient performed drawing tasks, the system modulated the stimulation levels appropriately: the system applied stimulation while the patient moved his arm and turned stimulation off while he was at rest. During the prompted movement task, the system responded during every action and rest period to enable and disable stimulation appropriately during the tasks. The system also performed well despite changes in beta-band power observed during periods when stimulation was turned on.

Due to considerable system delays, this system was not able to prevent tremor at the beginning of the patient's movement. Sensing packets are sent through the Nexus-D telemetry system every 400ms, and they take an additional 100ms to complete each transaction. Our signal processing method averages two seconds' worth of beta band-power readings before comparing that average to a pre-determined threshold. Once the threshold is exceeded, the system must wait another 400ms before the communication channel is free (while the system downloads the next packet of data), before finally beginning to ramp the stimulation up to the clinical value over the course of 1.2 seconds. As such, system delays alone are on the order of around three seconds.

There are a number of ways to reduce this delay in the future iterations of this system. In a commercial system, the signal processing and control computations can be moved onto the implanted device, removing the need for serialization through a small-bandwidth telemetry system. This would eliminate all communication and packet-interval delays, which currently contribute nearly 1 second of delay. More sophisticated classifiers might also provide faster and more robust recognition of movement and rest. These classifiers could potentially remove the need for beta-band power averaging and thresholding. However, given the expected variability in the neural biomarkers of movement across patients, automated classifier training methods will need to be used in order for these closed-loop systems to translate into general clinical practice.

It is important to consider what a final designed system's desired false-positive rate should be. Since existing systems stimulate constantly—at what can be thought of as a very high false-positive rate—any incremental reductions in stimulation time while maintaining tremor suppression is a clear improvement. By tuning the system to be more aggressive in applying stimulation, we may lose some specificity and have a higher false-positive rate, but obtain a reduced delay and thus do a better job of obtaining overall therapeutic benefit.

Another consideration is that other ECoG signals could be used for closed-loop control. For instance, prior work in the BCI field has identified that there are movement-related changes in both the beta and gamma bands [13,15]. While the system used in this experiment was not able to make use of the gamma band due to limitations imposed by communication data compression in the Nexus-D [9], gamma signals are more localized on the motor cortex and are more task specific than beta-band changes. Including gamma-band signals may improve a system's ability to determine when a patient is moving their limb and needs stimulation in the future. Additionally, research conducted with Parkinson's Disease patients suggests that some abnormal cortical signals may correspond to movement disorder symptoms [16]. Future work may find biomarkers that indicate when an ET patient is experiencing tremors, and such a biomarker would allow us to limit stimulation only to periods when the patient is experiencing tremor.

When discussing system performance, it is important to consider the patient's experience as modulation of stimulation parameters could have an impact on patient quality of life [8,17]. To gauge the potential impact of closed-loop control on the patient's everyday experience, we conducted semi-structured interviews with the patient during each research visit [18]. When asked if the tremor at the beginning of movement is bothersome, the patient responded, "It's not that big of a deal. You get used to it." Given that by the time patients receive a DBS system they have often adapted to living with tremor for quite some time, the transient effects may be of less concern to many patients, as long as they reach their desired tremor-free state in a predictable time frame.

However, it must be noted that in order for any algorithm to be viable, it must save more stimulation power than is used in the sensing and processing of data to deliver the closed-loop therapy. The exact cut-off for this balancing point would depend heavily on the hardware and firmware implementation, although with the continual drive to deliver lower-powered circuitry there may be a point where highly sophisticated algorithms may be used without concern of the sensing or processing power. It has been estimated that for an Activa PC+S system with typical stimulation parameters, a closed-loop DBS system that was triggered (or actively stimulating) for 94% of the time would have equal energy usage as an open-loop system [19]. This means that even minor or modest reductions in stimulation power will be worth the additional system components to allow devices to utilize closed-loop methods.

V. CONCLUSION

In conclusion, this paper documents the first successful demonstration in an ET Patient of a fully-implanted sensing and stimulation platform for neural-triggered, closed-loop DBS. To the best of the authors' knowledge, this is the first use of voluntary motor neural signals to drive therapeutic stimulation in real-time. Moreover, this technology, by providing access to simultaneous chronic recordings from deep brain and cortical structures, may enable increased understanding of the mechanisms of tremor and tremor suppression. While the current system uses an external computer for data logging, processing, and control, forthcoming versions of the IPG (from Medtronic) will likely be able perform these operations on hardware embedded within the implanted device, so that no external communication would be required. We have demonstrated here a system that is able to modulate delivery of DBS by identifying volitional movement, providing tremor reduction in an ET patient. Further work using this paradigm may realize a feasible, fully-implanted system for the delivery of demand-driven therapy for ET.